\documentclass[twocolumn,showpacs,preprintnumbers,amsmath,amssymb]{revtex4}

\usepackage{graphicx}% Include figure files
\usepackage{dcolumn}% Align table columns on decimal point
\usepackage{bm}% bold math

\begin{document} 

\title{Conductance in multiwall carbon nanotubes and semiconductor 
nanowires : evidence of a universal tunneling barrier.}
\author{J.-F. Dayen,  T. L. Wade,
M. Konczykowski, and J.-E. Wegrowe.}
\address{ Laboratoire des Solides Irradi\'es, Ecole 
Polytechnique, CNRS-UMR 7642 \& CEA/DSM/DRECAM, 91128 Palaiseau Cedex, France.}
\author{X. Hoffer}
\address{ Institut de Physique 
des Nanostructures, Ecole Polytechnique F\'ed\'erale de Lausanne, CH - 
1015 Lausanne, Switzerland.}

\begin{abstract}
Electronic transport in multiwall carbon nanotubes and
semiconductor nanowires was compared.  In both cases, the non
ohmic behavior of the conductance, the so-called zero bias anomaly,
shows a temperature dependence that scales with the voltage
dependence.  This robust scaling law describes the conductance
$G(V,T)$ by a single coefficient $\alpha$.  A universal behavior as a
function of $\alpha$ is found for all samples.  Magnetoconductance
measurements furthermore show that the conduction regime is weak
localization.  The observed behavior can be understood in terms of the
coulomb blockade theory, providing that a unique tunnel resistance on
the order of 2000 $\Omega$ and a Thouless energy of about 40 meV
exists for all samples.

\end{abstract}
 
\date{\today}

\maketitle

\newpage

There is intense interest in electronic transport in nanostructures in
various contexts, from single electron transistors to carbon
nanotubes, semiconductor nanowires, metallic nanoconstrictions or
other molecular structures.  In the presence of a tunnel junction, a
non-Ohmic behavior of the conductance $G$, termed zero-bias anomaly
(ZBA), is generally observed at low temperature.

For carbon nanotubes (CNT), the voltage dependence of the ZBA at low
temperatures and high bias is a power law $ G= G_{V} \cdot (eV)^{\alpha} $,
and the temperature dependence at low bias is also a power law, with
the {\it same power coefficient} $\alpha$; $ G= G_{T} \cdot (kT)^{\alpha} $
\cite{Bockrath,Schoenberg,Yi,Kanda,Hunger,Hoffer},
where $e$ is the electronic charge and $k$ the Boltzmann constant. 
Under this approximation, the conduction properties
$G_{\alpha}(kT,eV)$ can then be described for each sample (at zero
magnetic field) by a single scaling coefficient $\alpha$, and the two
prefactors $G_{V}$ and $G_{T}$.  Beyond this approximation, a more 
general description is given by a scaling function $f$, such that
$GT^{-\alpha} = f(eV/kT) $.

The scaling law is presented in the CNT literature as a 
manifestation of an underlying physical mechanism.  In the presence of
a tunnel junction, a Coulomb Blockade (CB) effect is expected.  In the
case of an ultrasmall junction, CB is described by the environmental impedance $
Z(\omega )$ \cite{Grabert1,Schoen}.  In more extended tunnel junctions
with disorder, the field and electrons propagate diffusively within
the electrodes, and non perturbative methods should be used.  Finally,
in the case of 1D systems, Luttinger Liquid states are expected.  In
all three cases, the conductance takes an identical form (see Eq. 
(\ref{GV}) below) under a rather general hypothesis
\cite{Grabert,Mishchenko,Egger}.  Furthermore,
measurements of the conductivity under applied magnetic field show
typical weak localization.  Accordingly, it is
possible to invoke either Luttinger liquid states
\cite{Bockrath,Schoenberg,Egger,Hunger,Zaitsev}, CB
effects \cite{Grabert1,Schoen,Kanda,Tarkiainen,Hoffer},
or diffusion effects related to disorder and weak localization
\cite{Schoenberg,Tarkiainen,Stojetz}.

In order to clarify the situation, the method followed in this work is
to correlate systematically the coefficient $ \alpha $ to other
experimental parameters, by performing a comparative study on a
statistical ensemble of samples.  In parallel with the CNT we present
a systematic study of the scaling law occurring in {\it semiconducting
nanowires} as a new argument in this debate.  

We measured two sets of samples. The first set of about 50 samples is
composed of nanotubes obtained by CVD on Ni or Co catalyst in a
nanoporous alumina membrane (the process is described elsewhere
\cite{Hoffer,Travis}).  The nanotubes are well separated (one nanotube
per pore) and are connected perpendicularly to a Au, Ni or Co contact. 
The diameter of the nanotube is calibrated by the diameter of the
pore.  One or a few nanotubes are contacted in parallel.  The
anodisation techniques allow the diameter of the pores to be well
controlled, from 40 nm down to 5 nm \cite{Hoffer,Travis}.  The
length of the nanotube (controlled by the length of the catalyst
electrodeposited inside the pores) was adjusted between about 1.5 $\mu$m
down to 100 nm.  The nanotubes are grown inside the pores by
standard CVD technique with acetylene at 640 $^{o}$C, after the
electrodeposition of Ni or Co catalyst.  The CNT are multiwalled.  The
top contact is made by sputtering, or evaporation, after the growth
of the tubes, and after exposing the samples to air. Different
materials and crystallinities have been used for the top contacts. 

The second set of samples is composed of tellerium semiconductors (Te)
obtained by electrodeposition in nanoporous polycarbonate or alumina
membranes \cite{Travis} of diameters $d$=40 nm and $d$=200 nm.  At 200
nm, the nanowires should no more be 1D with respect to electronic 
transport (because the energy 
separation between quantum levels $ \Delta E=(\pi
\hbar )^{2}/(2m^{*} d^{2})$, where $m^{*}$ is the effective mass, should
be above the thermal energy).  With the electrodeposition technique,
a single nanowire can be contacted {\it in situ} with a feed-back
loop on the intermembrane electric potential \cite{Travis}.  Both
contacts are free of oxides, due to the chemical reduction at the Te
interfaces during the electrodeposition, and due to in the {\it in
situ} contacts.  The Te are contacted with Au or Ni : $Au/Te/Au$ or
$Ni/Te/Ni$. 

The dynamical resistance measurements were performed with a lock-in
detection bridge LR7OO (using AC current of 0.3 nA for most
samples to 10 nA for low resistive samples) and a DC current. DC resistance
measurements were also made with a nanovoltmeter. The
temperature is ranged between 4K and 200K. A superconducting coil
gives a perpendicular magnetic field, ranging between $\pm$ 1.2 T. This
experimental protocol allows us to measure dynamical resistance as a
function of DC current amplitude, perpendicular magnetic fields, and
temperature.

The typical profile of the ZBA is plotted in Fig 1 for a Te
semiconducting wire (a and c) and for a CNT sample ((b) and (d)).  Two
well defined conductance regimes are observed : the voltage dependence
of the ZBA at low temperature and high bias $ G= G_{V} \cdot (eV)^{\alpha} $
and the temperature dependence at low bias, which is also a power law
with the same power coefficient $\alpha$; $ G= G_{T} \cdot (kT)^{\alpha} $. 
The ZBA vanishes above 50K, but the temperature dependence is also
valid at high temperature.  A more general description (which shows
the deviation to the simple power law approximation) is presented in
the form: $GT^{-\alpha} =f(eV/kT)$ (Fig 1 (c) and (d)).  A very
large majority of samples exhibit the scaling law (48 CNT over 55 with
enough length \cite{Hoffer} and 13 Te nanowires over 14).  This
scaling law is very robust since samples are different from the point
of view of the nature of the contacts, and the quality and
nature of the nanowires or nanotubes.  The CNT are contacted with Ni
or Co catalysts \cite{Travis} on the bottom.  The top of the wire is
contacted either with amorphous Ni, or with highly disordered Co
(mixed hcp and cfc nanocrystallites) or with single crystalline cfc Co
layer \cite{charact}.  The coefficient $\alpha$ for Co electrodes is
statistically larger than that for Ni (Fig 1 (f)).

\begin{figure}[h]
   \begin{center}
      \includegraphics[scale=0.3]{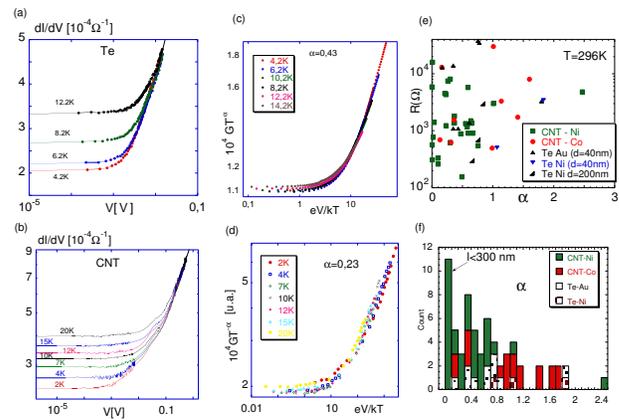}
   \end{center}
   \caption{Conductivity dI/dV as a function of bias voltage for
   different temperatures of a typical sample (a) Electrodeposited Te, (b) Carbon
   nanotube.  Scaling law of the quantity $G T^{-\alpha}$ for Te (c)
   and CNT (d).  (e) All samples: distribution of the resistances at room
   temperature as a function of $\alpha$ and (f) Histogram of scaling
   coefficient $\alpha $.}
\end{figure}

Most of the resistances at room temperature are distributed from about
300 to 40 000 $\Omega$ (Fig 1 (e)).  There are no statistical
correlations between the resistances at room temperature and the
coefficient $\alpha$.  The conductance variations from one sample to
the other are not due to the nature of the contact.  Fig 1 (f) shows
the corresponding histogram for $\alpha$.  The first peak near $\alpha
= 0$ is due to short CNTs, with a length $L \leq 300$ nm of the order
of the thermal length (i.e. the CNT are screened by the contacts).

There are no statistical correlations between the resistance and the
length or the diameter of the CNTs (not shown).  We define a ratio
$\eta = R(50K)/R(300K)$ as the resistance at 50K divided by the 
resistance at room temperature. The coefficient $\eta$ is a measure of the
contribution of the electrodes and interfaces at high temperatures
(i.e. with the exclusion of the contribution of the physical mechanism
responsible for the ZBA).  The parameter $\eta$ is correlated
to the coefficient $\alpha$ ( Fig 2 (a)), but the correlation depends
strongly on the nature of the electrodes.  A tendency is sketched by
the straight lines in Fig.  2 (a), and the most important deviation is
seen for CNT with single crystalline cfc Co.  This shows that
$\alpha$ is related to the contacts, and is not exclusively defined by
the states of the wires or tubes.  The coefficient $\alpha$ is hence
dependent on extrinsic parameters.  On the other hand, the coefficient
$\alpha$ is correlated to the length of the CNT (as shown in reference
\cite{Hoffer}), which indicates that $\alpha$ depends also on
intrinsic parameters (including defects).

However, the most important result of this study
is the unique relation existing between the prefactors $G_{V}$, $G_{T}$
and $\alpha$ (Fig 2 (b)), whatever the nature of the 
samples.  For each sample, the extrapolation at 1K
gives the conductance $G_{T}/k^{\alpha}$ (plotted
in Fig 2 (b)), and the extrapolation at 1V gives the coefficient
$G_{V}/e^{\alpha}$ (plotted in the inset of Fig 2 (b), same scale).  All points align
on the same curve.  

\begin{figure}[h]
   \begin{center}
      \includegraphics[scale=0.3]{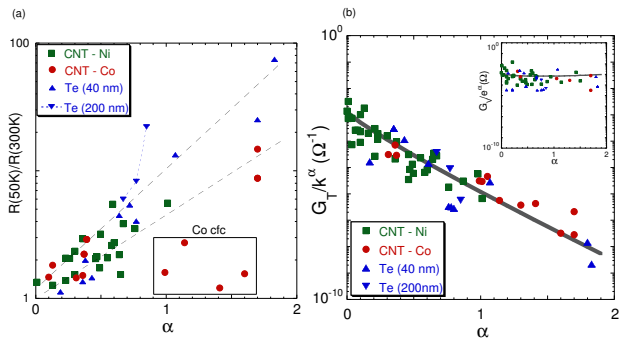}
   \end{center}
   \caption{(a) All samples : correlation between the ratio of the
   resistance at 50K and the resistance at 300K as a function of
   $\alpha$.  The lines are guides for the eyes.  (b) Conductivity
   $G_{T}/k^{\alpha}$ for all samples (extrapolated at T=1K) as a
   function of $\alpha$.  Inset : $G_{V}/e^{\alpha}$ (extrapolated at
   V=1 V).  The data are fitted from Eq.  (\ref{GV}) in the text, with
   parameters: resistance $R$ and energy $eV_{0}$ of a tunnel
   junction. }
\end{figure}

The correlation between $\alpha$ and the prefactors is $\it a \, \,  
priori$ not expected 
because there is no correlations between $G$ and $\alpha$ at room 
temperature. The function appearing 
in fig 2 (b) is a new universality exhibited by all measured samples, 
providing that the scaling law is measured. 
The discussion of the observed relation in terms of CB (curves 
fitted in Fig 2 (b)) follows.

In Fig 2(b), note that the difference between the fit in the main
figure and the fit in the inset is about $(e/k)^{\alpha} \approx 10^{4
\alpha}$, so that the two prefactors $G_{V}$ and $G_{T}$ are
approximately equal.  This means that the deviation from the
approximation of the function $G_{\alpha}(eV,kT)$ in the two power
laws is small even for intermediate regimes.

More information about the system, and especially about disorder and
quantum diffusion, can be obtained by applying a magnetic field $H$
perpendicular to the wire or tube axis
\cite{Tarkiainen,Schoenberg,Stojetz}.  Only the
magnetoconductance (MC) of CNTs (1.5 $\mu$m) and Te wires (about 5
$\mu$m) of fixed length are presented.  As plotted in Fig
3(a) a positive MC is present, but depends on the bias regime, low or
high.  At the high bias regime, the MC is destroyed and this effect is
not due to Joule heating, as seen in the Fig 3 (a) by comparing two
temperatures.  This observation is observed in all samples (including
semiconductor nanowires), and has not been reported previously.  In
the low bias regime, the MC exhibits all characteristics of weak
localization.  The MC curves at zero bias are fitted (Fig 3 (b))
with the 1D weak localization formula
\cite{Schoenberg,Tarkiainen,Stojetz} for :

\begin{equation}
 \Delta G_{WL}=-\frac{e^{2}}{\pi \hbar L} \left  ( l_{\Phi}^{-2} + 
 W^{2}/3l_{m}^{4} \right )^{-1/2}
 \label{WL}
 \end{equation}

where $l_{\Phi}$ is the coherence length, $l_{m} = \sqrt{\hbar / eB}$,
$ L$ is the length and $W$ is the radius of the wire.  The fit is
valid for all samples, except for the Te samples of diameter 200 nm 
(the large wires are no longer 1D with respect to the coherence length).  The
parameter $l_{\Phi}$, ranged between 50 and 300 nm, is greater than
the diameter of CNTs and wires, and follows the expected temperature
dependence $T^{-1/3}$ (inset of Fig 3 (b)).  The decrease in the
amplitude of MC with increases in the wire length and diameter has
been observed.  The presence of weak localization confirms the
diffusive nature of the transport, and confirms the high degree of
disorder.  The diffusion coefficient obtained with $l_{\Phi} \approx
100$nm is around $ D_{\Phi}¥ \approx 100$ cm$^{2}$/sec
\cite{Schoenberg,Stojetz}, confirming previous resuts about CNT.
However, the Fig 3 (c) shows that, surprisingly, the weak localization
is also strongly correlated to the coefficient $\alpha$.  In contrast
to the universal law plotted in Fig 2 (b), the relation between the MC
and the coefficient $\alpha$ depends on the nature of the contacts for
CNTs.  Two different curves are present for Ni and Co contacts to
CNTs.  A linear relation is observed for the Te of 40 nm diameter (the Au or Ni
electrodes cannot be differentiated).  Accordingly, $\alpha$ accounts
also for the diffusion mechanisms, and these mechanisms depend on the
nature of the interface. The coherence length is plotted as a 
function of $\alpha$ in the inset of Fig 3 (c).

\begin{figure}[h]
   \begin{center}
      \includegraphics[scale=0.3]{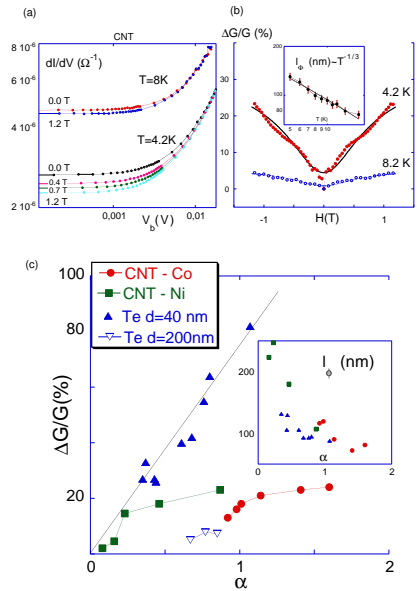}
   \end{center}
  \caption{(a) For a typical sample (here CNT - Co): magnetoconductance as a
  function of bias voltage for different magnetic fields at 4.2K and
  8K. The field is perpendicular to the wire.  (b) Magnetoconductance
  (same sample in \%) at zero bias for different temperatures fitted by Eq. 
  (\ref{WL}) to weak localization.  Inset : temperature dependence of
  the phase coherence length $l_{\Phi} \propto T^{-1/3}$.  (c) All samples :
  correlation between magnetoconductance and coefficient $\alpha$ at
  zero bias and 4.2K.}
\end{figure}

We now discuss the data in terms of CB theory. In the CB regime,
the coefficient $\alpha$ is defined by the action of the
electromagnetic environment on the charge carriers, or in terms of
transmission lines, by the impedance $Z$ of the circuit to which the
junction is contacted.  The coefficient $\alpha$ depends on the
diffusion constant of both the electromagnetic field and the charge
carriers (some expressions are given in \cite{Grabert,Egger}).  The
scaling is obtained if the spectral density of electromagnetic modes
$I(\omega)$ is finite at low energy down to zero frequency
modes : $\alpha =I(\omega \rightarrow 0)) = \frac{Z(\omega \rightarrow
    0)}{\left ( h/2e^{2} \right )}$.

The conductance at
zero temperatures (\cite{Grabert1}, Chap 2 formula (113) and
\cite{Grabert}, formula (19)), is given by Eq. (\ref{GV}) 
for the prefactor $G_{V}$ (below).  It has also been predicted that the
value at finite temperature and low bias coincides ( \cite{Schoen}
chap 3, p25 (3.63)) with the expression of $G_{V}$; the bias
voltage energy and the thermal energy $eV \leftrightarrow kT$ can be
permutated :

\begin{equation}
 G_{T} \approx G_{V} = \frac{1}{R} \, \frac{e^{- \gamma \alpha}
 }{\Gamma(2 + \alpha)} \left ( \frac{\pi \alpha}{eV_{0}} \right
 )^{\alpha}
 \label{GV}
\end{equation}
    
where $\gamma$=0.577 \ldots is the Euler constant and $\Gamma$ the
Gamma function.  The resistance of the tunnel barrier is $R$, and the
energy $eV_{0}$ is, in the case of ultra-small tunnel
junctions, the Coulomb energy $E_{C}=e^{2}/2C$, where $C$ is the
capacitance of the tunnel barrier.  In a diffusion regime, the
relevant energy is the Thouless energy $eV_{0} = E_{T}= \hbar
D_{T}/a^{2}$ where $D_{T}$ is the diffusion constant for the
electromagnetic field in the electrodes or for the charges, and $a$ the
relevant length (the capacitance is now included in the coefficient
$\alpha$) \cite{Grabert,Pierre}. 

 As already mentioned, the power law is observed in Fig 1 (a) and (c). 
 However, it is very surprising that Eq.  (\ref{GV}) also fits the
 data plotted in Fig 2 (b) as a function of the coefficient $\alpha$. 
 The only fitting parameters are now the tunnel resistance $R$ and the
 energy $eV_{0}$.  This means that {\it all samples have the same
 tunnel barrier} (within the tolerance of one order of
 magnitude over nine).  The fit with Eq.  (\ref{GV}) of the data
 $G_{T}$ plotted as a function of $\alpha$ (in Fig 2 (b), after
 correction of the ratio $1/k^{\alpha}$) gives a tunnel resistance on
 the order of $R$=2.5 k$\Omega$, and an energy of about 40 meV (which
 corresponds to a capacitance of about $C$= 2 10$^{-18}$ F).  The fit
 of the data $G_{V}$ is less convincing, but
 gives, however, the same tunnel resistance and an energy
 of about 100 meV.  The relation $G_{V} \approx G_{T}$ is confirmed
 within the approximation of a scaling function $f$ composed of two
 power laws. For a typical length of a few
 nanometers, the diffusion constant coincides with the diffusion obtained
 from the weak localization $D_{T} \approx D_{\Phi} \approx 100$
 cm$^{2}$/sec.  In contrast, the diffusion constant $D_{T}$ deduced
 from the coherent length $a=l_{\Phi}$ is about $D_{T}=1000$ 
 cm$^{2}$/sec.  This value is one order of magnitude larger than
 $D_{\Phi}$ measured under magnetic field.

Why, despite the huge dispersion of intrinsic and extrinsic parameters
(and especially the typical sizes and disorder of the electrode/wire
junction), the parameters of the "tunneling junction" deduced from the
scaling law are universal \cite{Schottky}?  In other terms, why is it
{\it not possible} to differentiate between the samples from the point of view
of the CB? These results suggest that disorder and quantum diffusion,
together with relatively low dimensionality, impose a universal value
to the relevant Thouless energy and resistance involved. The origin
of this universality is not known.

In conclusion, a comparative study of electronic transport between
multiwall carbon nanotubes and Te nanowires has been performed.  The
samples are defined by a single scaling coefficient $\alpha$
describing the ZBA. A universal relation is observed beween $\alpha$
and the conductance, valid whatever the nature of the electrodes, the
lengths ($\mu$m range), and the diameters, ranged between 5 to 200 nm. 
All samples, except the 200 nm diameter Te, exhibit a typical 1D weak
localization behavior from which the coefficient $\alpha$ is also
correlated. This study shows that the scaling law of the ZBA
originates from a quantum diffusive process together with coulomb
blockade with a universal tunnel barrier. An interpretation of the
scaling law in terms of Luttinger liquid states can hardly be
maitained.

This work was partially suported by Swiss NSF and French DGA. We thank
H. Bouchiat, S. Roche, G. Montambaux, A. Bachtold, H. Pothier, and G.
L. Ingold for valuable discussions, and D. Pribat, J.-M. Padovani and
C. S. Cojocaru for their help in the CVD growth of 5 nm diameter CNTs.


\begin{references}
\bibitem{charact} The crystallinity of the contacts has been 
characterized by XRD and SQUID magnetometry.
\bibitem{Schottky} A metal-semiconductor junction would lead to a
Schottky barier, which can be measured in the
thermoemission regime.  However, since the power law
holds at high temperatures, it is not compatible with an activation
regime.
\end{references}
\end{document}